\documentclass[preprint,showpacs,preprintnumbers,amsmath,amssymb]{revtex4}

\usepackage{graphicx}
\usepackage{dcolumn}
\usepackage{bm}
\usepackage{epsfig}
\usepackage{subfigure}

\begin{document}

\preprint{APS/123-QED}

\title{Band Structure of Honeycomb Photonic Crystal Slabs}

\author{Tai-I Weng}
\author{G. Y. Guo}%
\email{gyguo@phys.ntu.edu.tw}
\affiliation{
Department of Physics, National Taiwan University, Taipei, Taiwan 106, Republic of China
}

\date{\today}

\begin{abstract}
Two-dimensional (2D) honeycomb photonic crystals with cylinders and connecting walls have the 
potential to have a large full band gap. In experiments, 2D photonic crystals do not have 
an infinite height, and therefore, we investigate the effects of the thickness of the walls, 
the height of the slabs and the type of the substrates on the photonic bands and gap maps of 
2D honeycomb photonic crystal slabs. The band structures are calculated by the plane wave expansion 
method and the supercell approach. We find that the slab thickness is a key parameter affecting 
the band gap size while on the other hand the wall thickness hardly affact the gap size.
For symmetric photonic crystal slabs with lower dielectric claddings, the height of the slabs needs to 
be sufficiently large to maintain a band gap. For asymmetric claddings, the projected band diagrams 
are similar to that of symmetric slabs as long as the dielectric constants of the claddings do not
differ greatly.

\end{abstract}

\maketitle

\section{Introduction}

Photonic crystals have been a major research field for scientists and engineers, for their 
capabilities of controlling light propagation \cite{photonic_crystals}. The pursuit of photonic 
band gap has been a major topic in studying photonic band 
structure \cite{photonic_band_gaps_and_localization} because many applications of photonic crystals 
are based on photonic band gaps. In the band gap region, there are no optical modes and spontaneous 
emission. One can control the light propagation direction in the band gap more easily. That is why 
optimizing the size of band gap is an interesting issue.   

\begin{figure}
\graphicspath{{images/}}
\includegraphics[width=6cm]{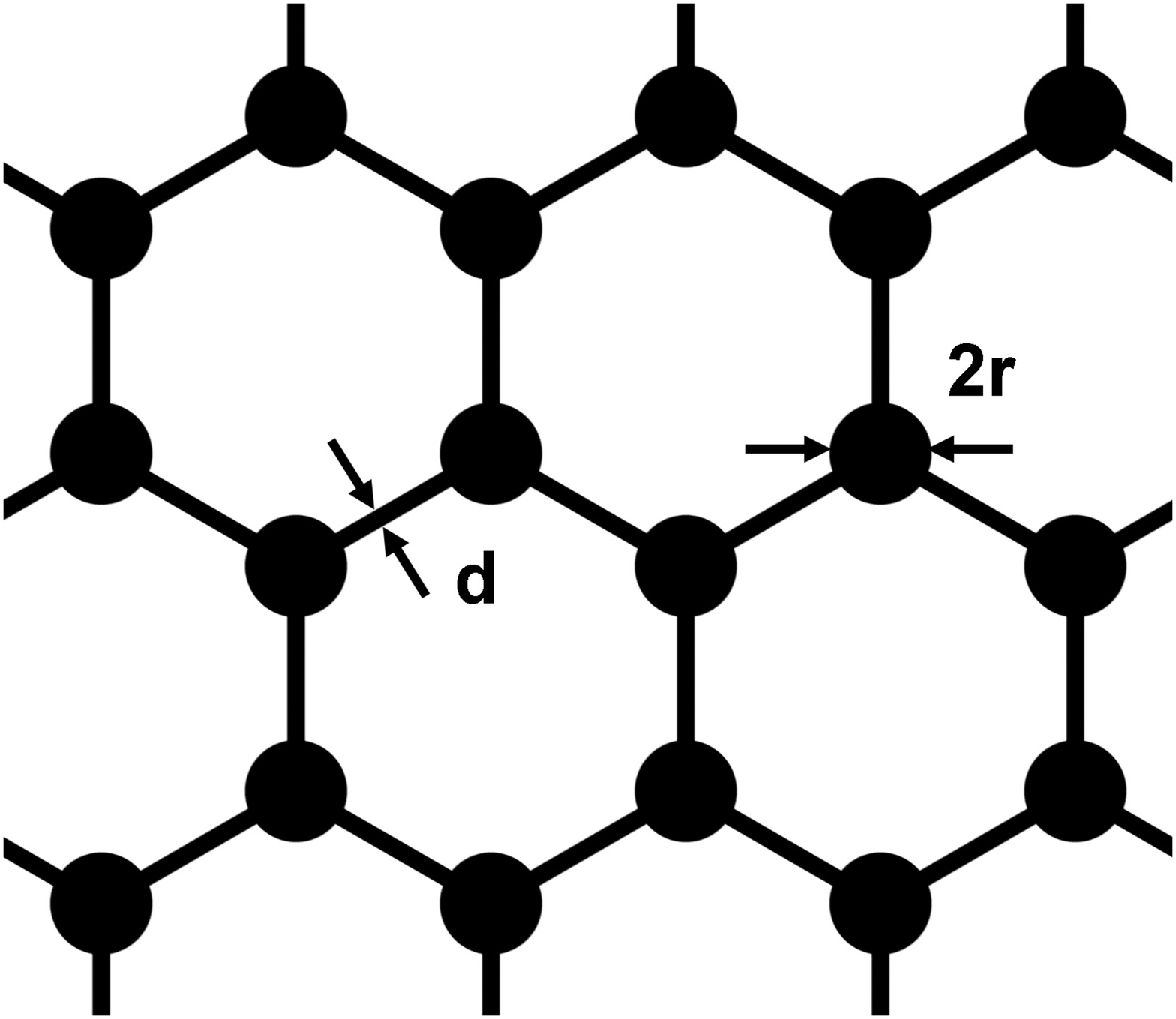}
\caption{ \\}
\vspace{4cm}
\label{fig:honeycomb}
\end{figure}

Chern {\it et al.} have recently proposed a two-dimensional (2D) honeycomb photonic crystal 
structure \cite{pre68_026704}
(Fig. \ref{fig:honeycomb}). The two-dimensional honeycomb photonic crystal
without the walls and also its slab structure was reported earlier, in 
Refs. \cite{photonic_crystals} and \cite{fan97}, respectively. 
There are two geometrical parameters in the two-dimensional honeycomb 
photonic crystal \cite{pre68_026704}: the radius of the 
cylinders, and the thickness of the walls. The reason to choose such a geometry is that the 
transverse magnetic (TM) type band gaps are favored in isolated high dielectric constant 
($\varepsilon$) region, 
and the transverse electric (TE) type band gaps are favored in connected 
lattice \cite{photonic_crystals}. The cylinders are isolated dielectric medium units, and the walls 
connect them. The full band gap can be optimized if we strike a balance between these two 
characteristics. Chern {\it et al.} \cite{pre68_026704} 
reported a largest full band gap of two-dimensional photonic 
crystals in the literature. Very recently, Fu {\it et al.} \cite{yfchen} have done some experiments on honeycomb 
photonic crystals. They found that the gap frequency and the gap-midgap ratio do not agree very 
well with the theoretical work of Chern {\it et al.} The underlying reasons still need further 
investigations.

Based on their studies, we have done some further calculations. The technique Chern {\it et al.} used 
is finite difference multigrid method. We use plane wave expansion method as implemented in the MPB
package \cite{op8_173}. Fully-vectorial eigenmodes of Maxwell's equations with periodic boundary 
conditions were computed by conjugate-gradient minimization of the block Rayleigh quotient in a 
plane wave basis \cite{op8_173}. Since in experiments the photonic crystals do not have an infinite 
height, we investigate not only two-dimensional structures but also photonic 
crystal slabs. The hexagonal lattice slabs without the walls have already been studied 
before \cite{fan97,prb60_5751}. In our slab case we analyze the effects of the thickness of the 
walls and the height of the slabs. We also discuss the substrate impacts. 

This paper is organized as follows. First we will introduce the theoretical formulas about the theory 
and techniques used in our calculations, including the plane-wave method, supercell technique and 
the projected band diagram for photonic crystal slab calculations in the next section. 
Then we will present the results of our calculations for two-dimensional honeycomb photonic crystals, 
and honeycomb photonic crystal slabs in Sec. III. For two-dimensional systems, we will show the band 
structures and gap maps. For slab systems, we will show the band structures and discuss the geometric 
and substrate effects. Finally we will make a brief summary of our work in Sec. IV.


\section{Computational Method}


The band structures are calculated by plane wave expansion method, a frequency-domain method which expands 
the fields in the plane-wave basis, directly solves the eigenstates and eigenvalues of Maxwell's 
equations, as implemented in the MPB package \cite{op8_173}. 
To analyze in an easier but exact way, we assume that light propagates in 
linear, time-invariant, lossless, magnetic uniform material ($\mu=1$). 

The slabs are not perfect three-dimensional photonic crystals since there is no periodicity in 
the $z$ direction. We use the supercell method, which introduce defects periodically. Since the 
slabs are periodic only in the $x-y$ plane, we add the original finite height cell with a sufficient 
amount of background region in the $z$ direction. Now it is a three-dimensional case and takes 
longer calculation time. If the background region is large enough and the light is confined 
in the central region of the cell and far enough from the borders, the boundaries will not affect 
the result too much. In other words, this technique is very useful for the modes confined in the slab \cite{roadmap}. 


For photonic crystal slabs, there is only two-dimensional periodicity in the $x-y$ plane, and the 
wave vectors are conserved in that plane, too. Since the wave vector in the $z$ direction is not 
conserved, only the projected band diagrams on the plane will be plotted. That is, although we are 
investigating a three-dimensional system, our band structure only involves the $k$-points in 
the $x-y$ plane.

The lightcone is an important feature of the projected band diagram. It is determined by the equation 
\begin{equation}
	\frac{\omega}{c} > \frac{k}{n_c}
\label{eqn:49}
\end{equation}
where $n_c$ is the refractive index of the cladding. Eqn. \ref{eqn:49} comes from the concept of total internal reflection. In the area below the lightcone, light propagates within the slab. In the area in the lightcone, light propagates outside the slab with a radiation loss. 

In two dimensions, we always decompose the electromagnetic modes into two noninteracting modes: TE 
(polarization of electric field confined in the plane) and TM (polarization of magnetic field 
confined in the plane) mode. In slabs the modes are not purely TE or TM modes, but they can still 
be classified as vertically even or odd modes with respect to the horizontal symmetry plane bisecting 
the slab. The $H_z$ component has a symmetrical field distribution for even modes and an asymmetrical 
field distribution for odd modes. Besides, for the first-order modes which have no node in the 
vertical direction, the field distributions within the core are very similar to the corresponding 
modes in infinite 2D photonic crystals. Moreover, in the mirror plane, the modes are purely TE 
(or TM) polarized. Therefore we can roughly regard even modes as being TE-like and odd modes as 
being TM-like \cite{pn1_1}.

\section{Results and Discussion}

\subsection{Two-dimensional Honeycomb Photonic Crystals}


Fig. \ref{fig:2dband} is the band structure of 2-D honeycomb photonic crystals, calculated by 
plane wave expansion method. This figure agrees well with the results calculated by multigrid method 
(Fig. $4$ in \cite{pre68_026704}). Here normalized frequencies ($f' = \omega a/2\pi c$) and 
wave vectors are applied. From this figure we can see that for both TE and TM modes, no light 
within $f'=0.388$ to $f'=0.492$ are allowed in this structure. This means that there is a complete 
band gap in this system. We define a gap-midgap ratio (ratio between band gap width and midgap 
frequency) to measure how large the photonic band gap is. Even if the scale of the system is changed, 
this quantity remains the same.


We plot a gap map by plane-wave method to see the effect of the wall thickness. 
Fig. \ref{fig:2dgapmap} is the relationship between gap-midgap ratio and wall width $d$ 
(with fixed $r/a=0.155$).  First we notice that as $d/a$ increases, the frequencies decrease. 
This is because of larger dielectric fraction and average index \cite{photonic_crystals}.
Moreover, in some frequency ranges, the TE mode gaps and the TM mode gaps overlap with each 
other and form complete band gaps. Near $d/a=0.035$ the complete band gap is the largest. 
From gap map we can find out the optimal wall thickness easily.

\subsection{Geometric Effects on Honeycomb Photonic Crystal Slabs}

Fig. 4 shows the band structure of air-bridged honeycomb photonic crystal slabs. 
The slab thickness is one of the key parameters in determining the band gap size in photonic crystal 
slabs. For too thin slabs, the slabs do not provide sufficiently strong perturbation to the 
background. The modes can propagate outside the slabs easily and therefore the guided modes will 
be very close to the light cone. For too-thick slabs, higher-order modes with horizontal nodal 
planes lie slightly above the lowest-order mode beause of a little more energy. Therefore no gaps 
exist \cite{prb60_5751}. Fig. \ref{fig:gapsize_h} shows the gap size as a function of slab thickness. 
The gap-midgap ratio is optimized when the slab thickness is equal to 0.8 $a$.



Fig. \ref{fig:gapsize_d} is the gap size figure of honeycomb photonic crystal slabs, and 
Fig. \ref{fig:gap_d} is the gap map of various wall thickness of slab structures. It is noticeable 
that when the wall thickness is larger than $d/a=0.13$, the gapsize remains the same while the 
frequency range of gap differs. We know that the key parameter affecting band gap size is the slab 
thickness, not the wall thickness. This reveals that the tolerance of wall thickness fabrication can 
be very large. Band gap size is not sensitive to wall thickness. With the same gap size, we can 
obtain different band gap frequency by tuning the wall thickness.

%

\subsection{Substrate Effects on Honeycomb Photonic Crystal Slabs}

Both symmetric and asymmetric triangular lattice of circular air cylinders in dielectric slabs have 
been studied before \cite{ijqe38_891,prb60_5751,prb66_033103}. Here we focus on honeycomb photonic 
crystal slabs. We consider two examples for symmetric photonic crystal slabs: air-bridged photonic 
crystal slabs and weak-confinement symmetric photonic crystal slabs. 


Air-bridged photonic structures consist of a thin 2-D photonic crystal in a high-index membrane 
(for example, silicon) surrounded by air (Fig. \ref{fig:symslaba}). Here the index contrast between 
the core and the cladding is very high so the light is strongly confined in the slab. However, 
it is not easy to integrate into a chip and this is the major disadvantage of this 
system \cite{ijqe38_743}.


Fig. \ref{fig:symslabb} is an example of weak-confinement symmetric photonic crystal slabs. It 
consists of silicon core and silicon dioxide claddings. If we compare Fig. \ref{fig:sio2band} to 
Fig. \ref{fig:slabband}, we can see that if we want to obtain the same order of band gap size, the 
height of the structured slab should be larger. This is because with the same height of 
air-bridged structures, the perturbation photonic crystal provided is not strong enough, and we will 
not obtain any band gap in this case. On the other hand, we should provide more perturbation to the 
background by adding the height of the slabs.


As for the asymmetric photonic crystal slabs, we consider the silicon-on-insulator (SOI) systems. 
In a SOI system, the lower cladding 
usually consists of an oxide layer, and the upper cladding consists only of air 
(Fig. \ref{fig:airsio2}). This structure is easier to integrate onto a chip than a membrane, but the 
asymmetry of the structure leads to additional losses when $\varepsilon$ of one of the claddings 
is lower.  

Since the slab is not symmetric anymore, the modes can not be identified as pure even or odd modes. 
But if we compare Fig. \ref{fig:airsio2band} to Fig. \ref{fig:sio2band}, we can see that 
although there are some slight differences, we can still distinguish them as even-like or 
odd-like modes. 
The reason is that the $\varepsilon/\varepsilon_0$ of upper cladding does not vary a lot in these 
two cases. Therefore the modes do not vary very much in the symmetric (Fig. \ref{fig:symslabb}) 
and asymmetric (Fig. \ref{fig:airsio2}) structures. The SOI system behaves in a way similar to 
the asymmetric one.


\section{Conclusions}

Tuning the frequency range of photonic crystal band gap is a very important issue in photonic crystal 
applications. We have done further studies of the honeycomb photonic crystal that Chern {\it et al.} 
proposed in 2003. This structure has the potential of a large complete photonic band gap. Two geometric
parameters, namely, the radius of the cylinders and the thickness of the walls, provide more 
flexibilities of designing photonic crystals. 
For two-dimensional honeycomb photonic crystals, we have calculated 
the gap map for both TM and TE mode and 
optimize the complete band gap. The band structure tells us that the complete band gap is 
large, as reported before \cite{pre68_026704}.

The main topic of this paper is, however, the photonic bands and gap maps of honeycomb photonic 
crystal slabs. For honeycomb photonic crystal slabs, projected band diagrams have been calculated. 
We focus 
on the guided modes below the light cone boundary. We find that the band gaps of two 
independent polarizations do not overlap with each other anymore. Our results show that the slab 
thickness is the key parameter of the band gap size and we have also determined the optimal slab 
thickness. On the other hand, the wall thickness does not affect the gap size very much. We obtain 
a saturated gap size when the wall thickness is above a certain value. This means that the tolerance 
for slab thickness fabrication is small, but large for wall thickness. We have also discussed about 
the effects of slab cladding, including the symmetric and asymmetric ones. For symmetric claddings, 
those slabs with weak confinement claddings should be higher to provide stronger perturbation to 
the background. For asymmetric claddings, as long as the dielectric constants do not differ a lot, 
the guided modes are much like the modes of symmetric slabs. We can use the asymmetric slab 
structures instead of the symmetric ones since asymmetric structures are more commercially 
available.

In short, this work reveals various properties of honeycomb photonic crystal slabss. Hopefully,
the optimized parameters would provide useful tips for experimentalists.

\begin{acknowledgments}

We thank Yang-Fang Chen for bringing their experimental work on 2-D honeycomb photonic 
crystals \cite{yfchen} to our attention, and Ken-Ming Lin for helpful discusions. 
We also gratefully acknowledge financial supports from National Science Council,
Ministry of Economic Affairs (94-EC-17-A-08-S1-0006) and
NCTS/TPE of the Republic of China.
\end{acknowledgments}

\newpage
\noindent {\bf Figure captions}\\

\noindent Fig. 1 The honeycomb structure: $r$ is the radius of the cylinders; $d$ is the thickness of the walls connecting the cylinders. \\

\noindent Fig. 2 Band structure for
two-dimensional honeycomb photonic crystals with $\varepsilon/\varepsilon_0=13$, $r/a=0.155$, and $d/a=0.035$. The horizontal dotted lines denote the band gap region. The normalized
frequency $f' = \omega a/2\pi c$. \\

\noindent Fig. 3 Gap map for two-dimensional
honeycomb photonic crystals with $\varepsilon/\varepsilon_0=13$ and $r/a=0.155$. \\
\noindent Fig. 4 Band structure for
air-bridged honeycomb photonic slabs with $\varepsilon/\varepsilon_0=11.9$, $r/a=0.155$, $d/a=0.035$
and $h/a=0.4$. \\
\noindent Fig. 5 Even mode gap size (as the percentage of
midgap frequency) versus slab thickness of air-bridged honeycomb photonic slabs
with $\varepsilon/\varepsilon_0=13$, $r/a=0.155$ and $d/a=0.035$.\\
\noindent Fig. 6 Even mode gap size (as the percentage of midgap
frequency) versus wall thickness for air-bridged honeycomb photonic slabs 
with $\varepsilon/\varepsilon_0=13$, $h/a=0.8$ and $r/a=0.155$. \\
\noindent Fig. 7 Even mode gap map
versus wall thickness for air-bridged honeycomb photonic crystal slabs
with $\varepsilon/\varepsilon_0=13$, $h/a=0.8$ and $r/a=0.155$. \\
\noindent Fig. 8 Side view of symmetric photonic crystal slabs with 
$\varepsilon/\varepsilon_0=1$ for air, $\varepsilon/\varepsilon_0=2.1$ for SiO$_2$.\\
\noindent Fig. 9 Band structure of weak-confinement symmetric honeycomb photonic crystal slabs with
 $h/a=2.0$, $\varepsilon/\varepsilon_0=1$ for air, $\varepsilon/\varepsilon_0=2.1$
for SiO$_2$, $\varepsilon/\varepsilon_0=11.9$ for silicon. \\
\noindent Fig. 10 Side view of a silicon-on-insulator system with
$\varepsilon/\varepsilon_0=1$ for air, $\varepsilon/\varepsilon_0=2.1$ for SiO$_2$.\\
\noindent Fig. 11 The band structure of asymmetric honeycomb photonic crystal slabs 
with $h/a=2.0$, $\varepsilon/\varepsilon_0=1$ for air, 
$\varepsilon/\varepsilon_0=2.1$ for SiO$_2$, $\varepsilon/\varepsilon_0=11.9$ 
for silicon. \\

\newpage

\begin{figure}[h]
\begin{center}
        \centering
        \graphicspath{{images/}}
        \includegraphics[width=8cm]{TIWengFig2.eps}
\end{center}
\caption{ }
\label{fig:2dband}
\end{figure}

\begin{figure}[h]
\begin{center}
        \centering
        \graphicspath{{images/}}
        \epsfig{file=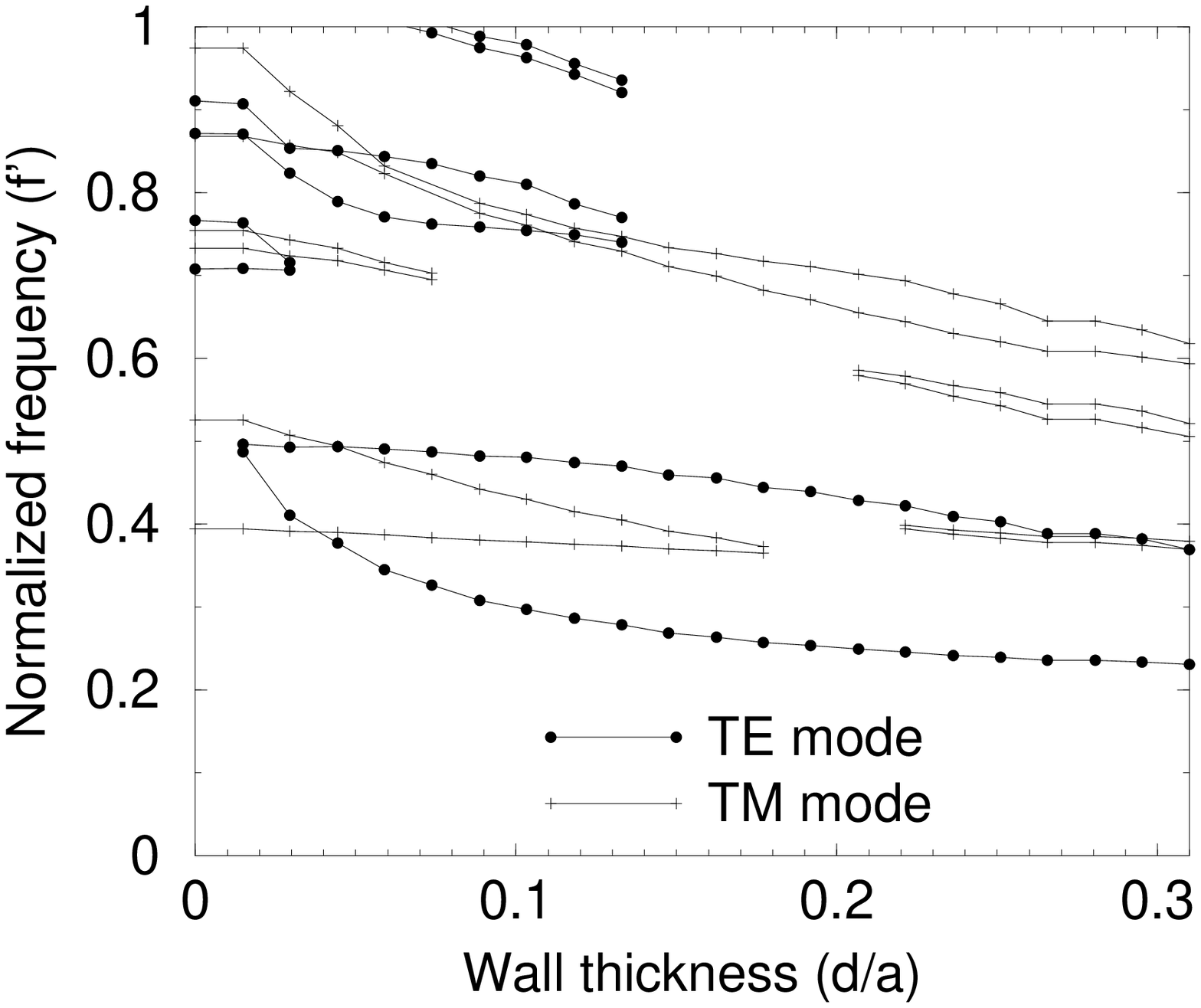,width=8cm}
\end{center}
\caption{ }
\label{fig:2dgapmap}
\end{figure}

\begin{figure}[h]
\begin{center}
        \centering
        \graphicspath{{images/}}
        \epsfig{file=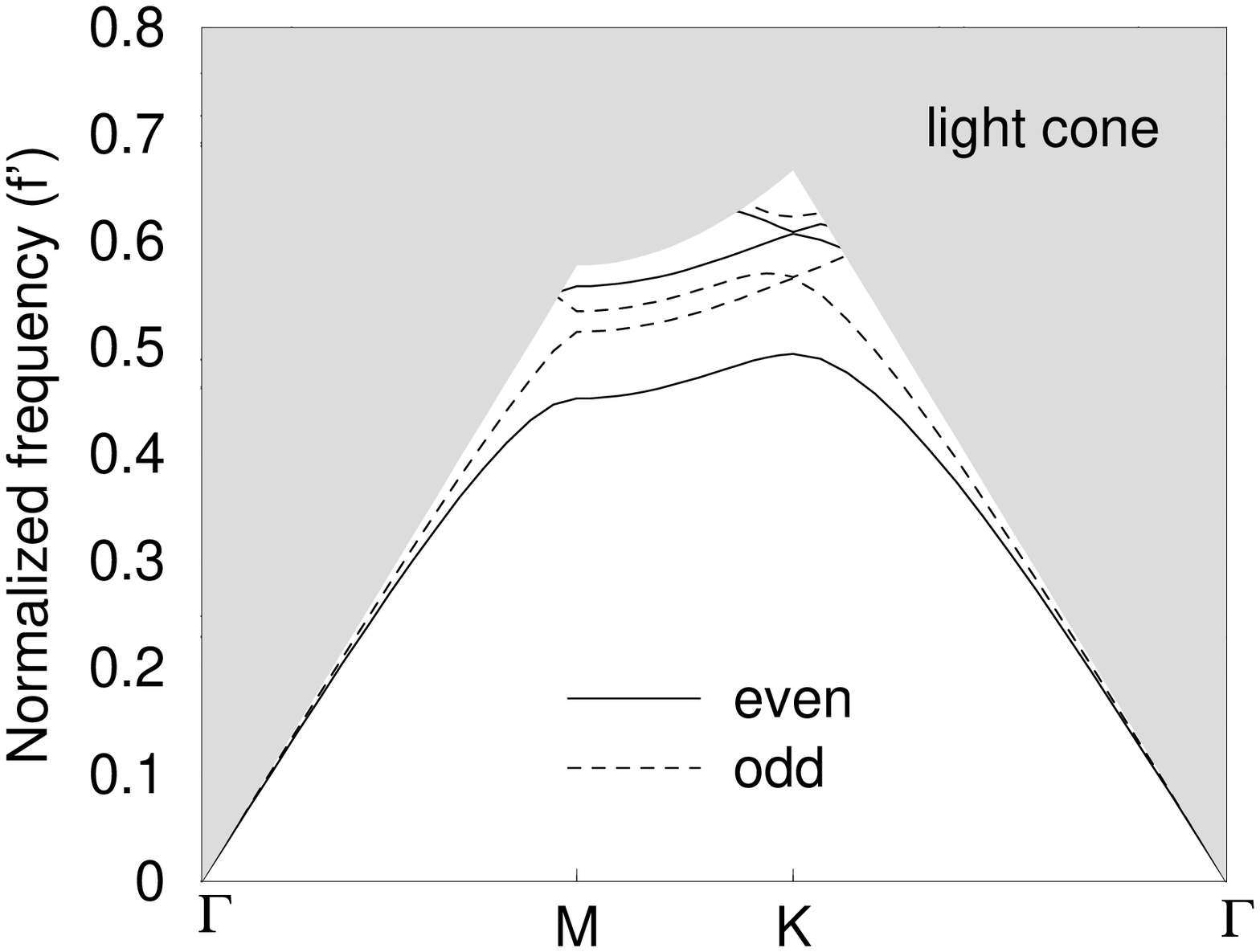,width=8cm}
\end{center}
\caption{ }
\label{fig:slabband}
\end{figure}

\begin{figure}[h]
\centering
\graphicspath{{images/}}
\epsfig{file=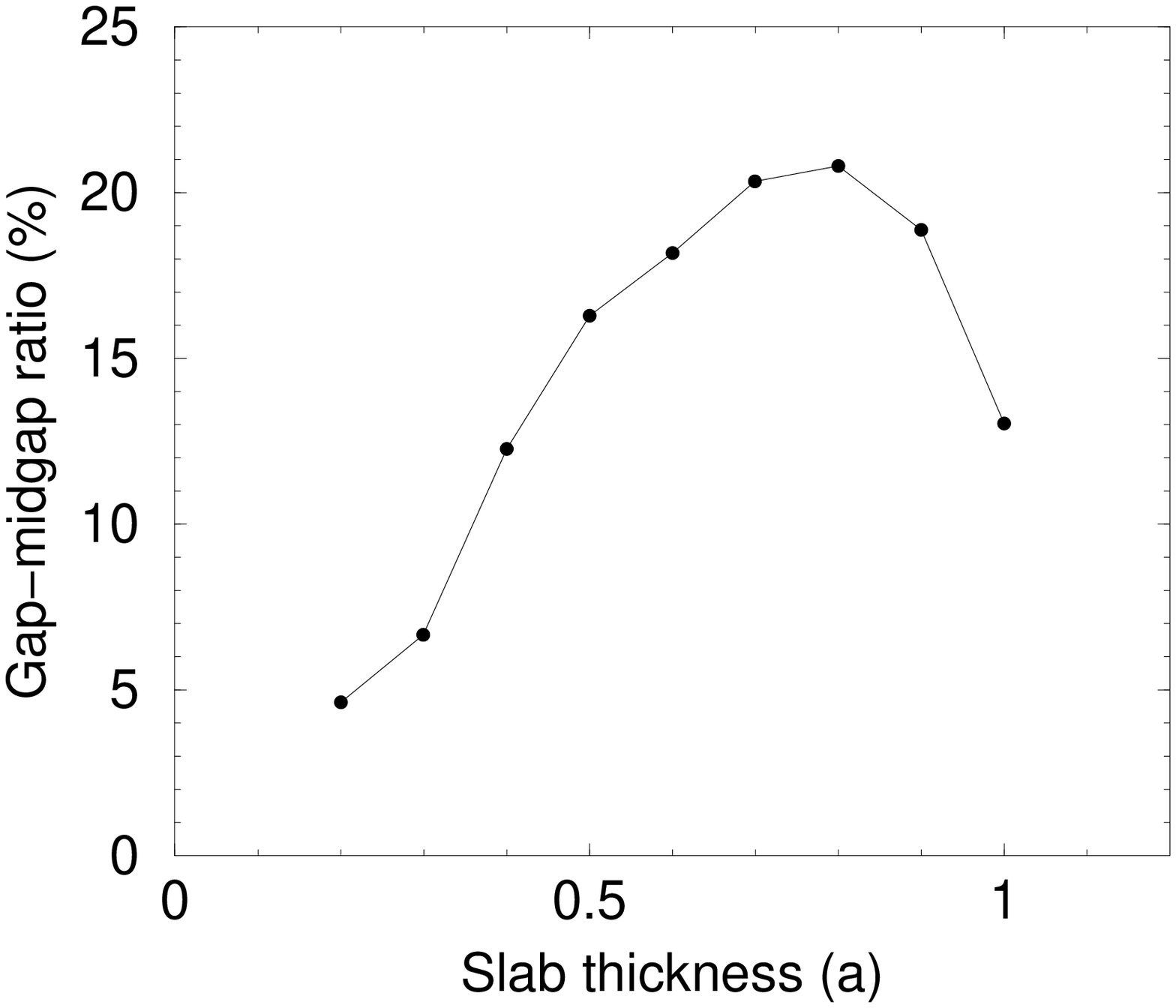,width=8cm}
\caption{ }
\label{fig:gapsize_h}
\end{figure}

\begin{figure}[h]
\centering
\graphicspath{{images/}}
\epsfig{file=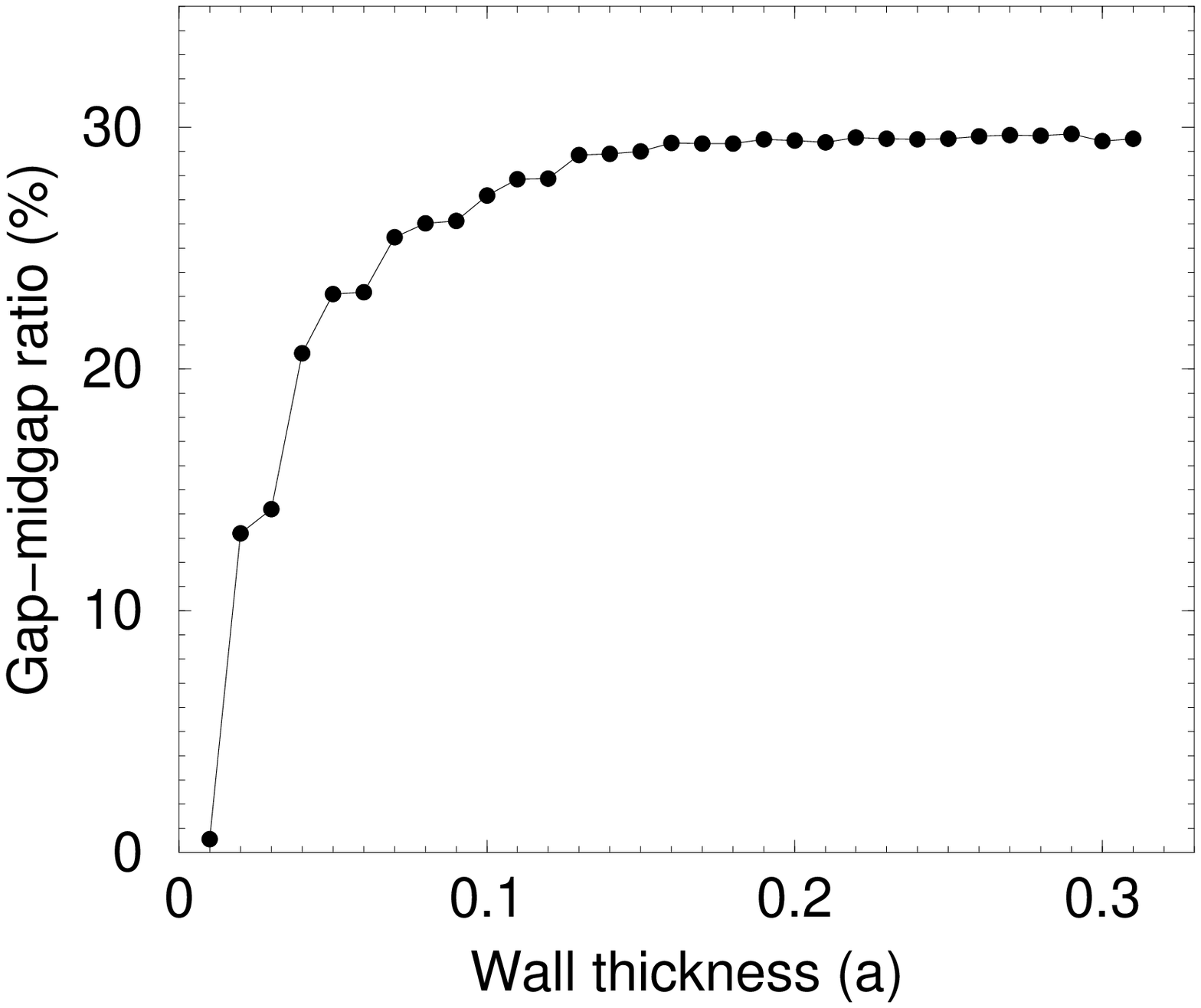,width=8cm}
\caption{ }
\label{fig:gapsize_d}
\end{figure}

\begin{figure}[h]
\centering
\graphicspath{{images/}}
\epsfig{file=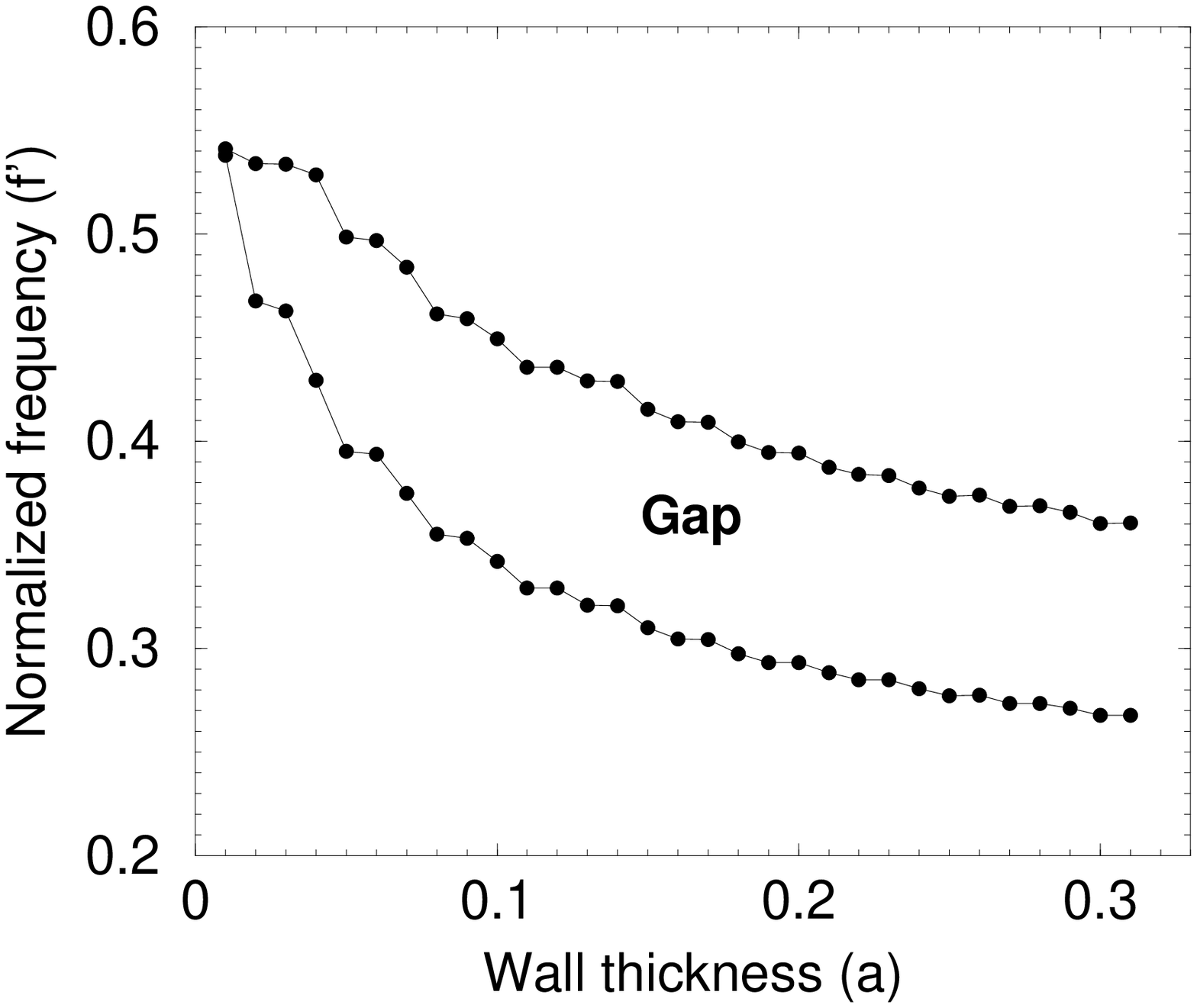,width=8cm}
\caption{ }
\label{fig:gap_d}
\end{figure}

\begin{figure}[h]
        \centering
        \graphicspath{{images/}}
        \subfigure[Air-bridged structure]{\epsfig{file=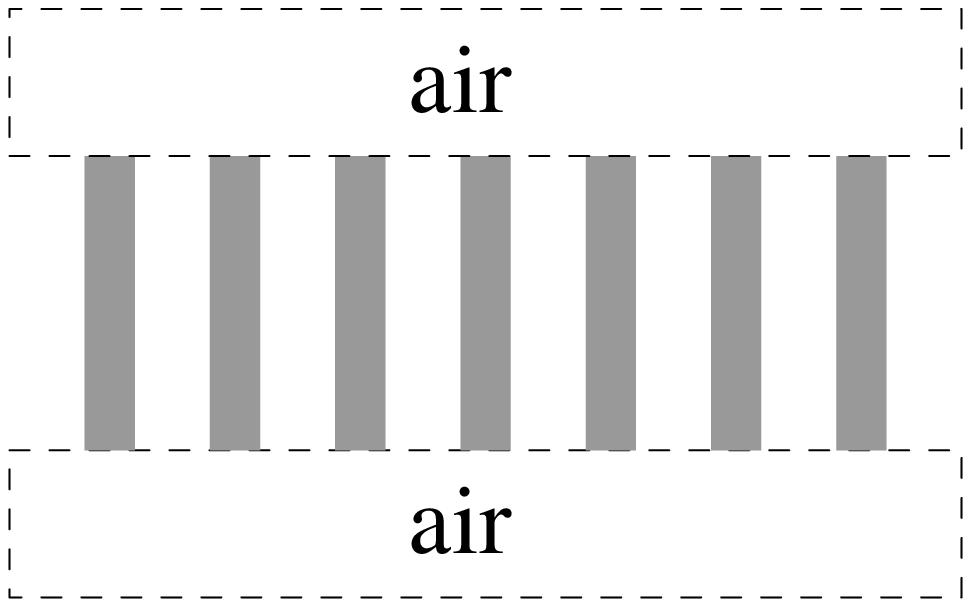,width=4cm} \label{fig:symslaba}}
        \subfigure[SiO$_2$ background]{\epsfig{file=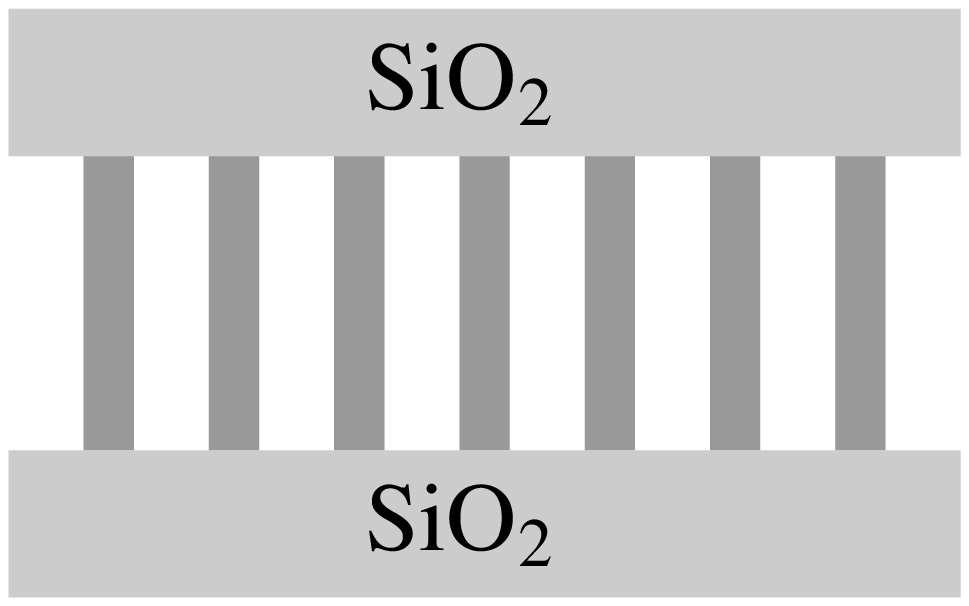,width=4cm} \label{fig:symslabb}}
\caption{ }
\end{figure}

\begin{figure}[h]
\begin{center}
        \centering
        \graphicspath{{images/}}
        \epsfig{file=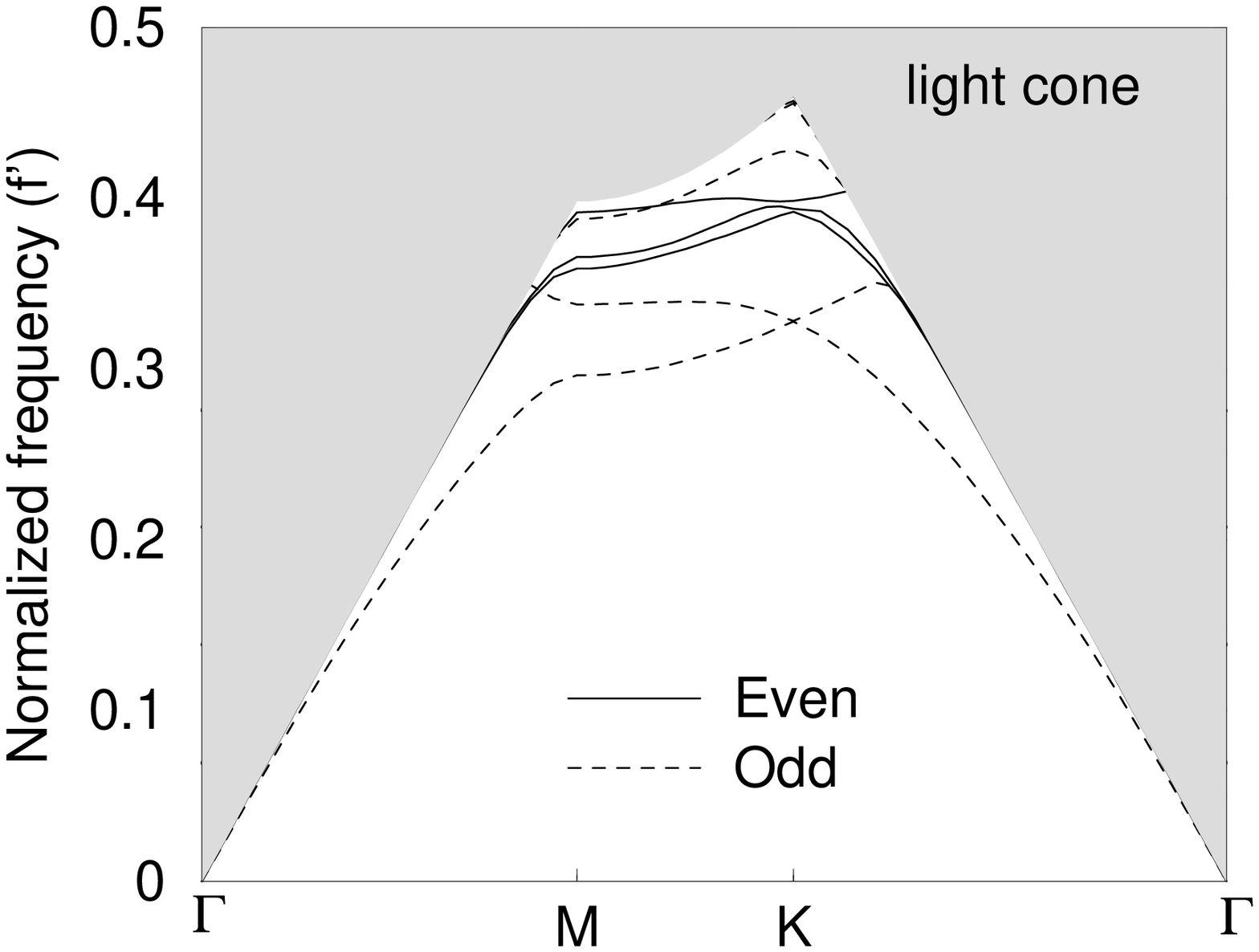,width=8cm}
\end{center}
\caption{ }
\label{fig:sio2band}
\end{figure}

\begin{figure}[h]
\begin{center}
        \centering
        \graphicspath{{images/}}
        \epsfig{file=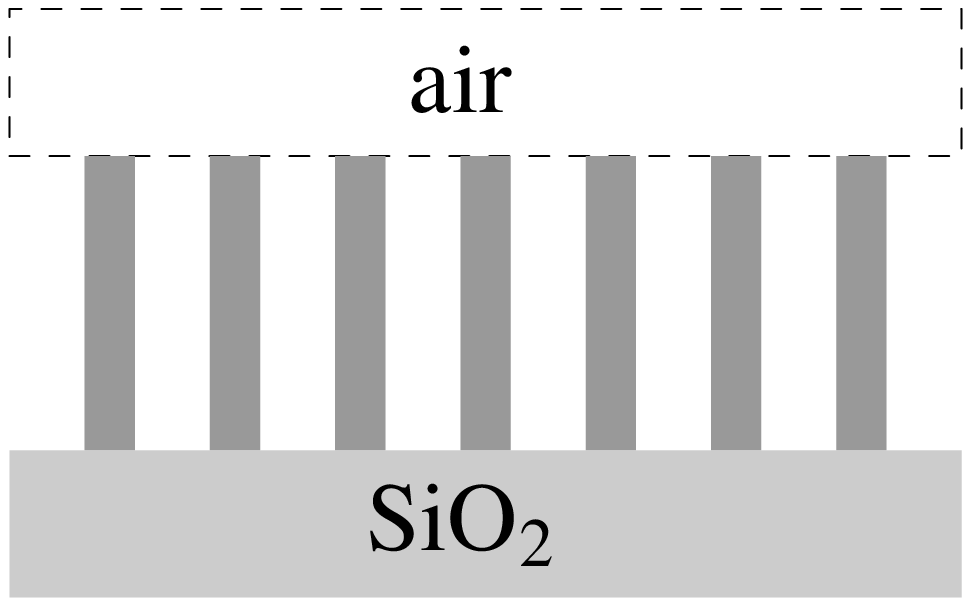,width=4cm}
\end{center}
\caption{ }
\label{fig:airsio2}
\end{figure}

\begin{figure}[h]
\begin{center}
        \centering
        \graphicspath{{images/}}
        \epsfig{file=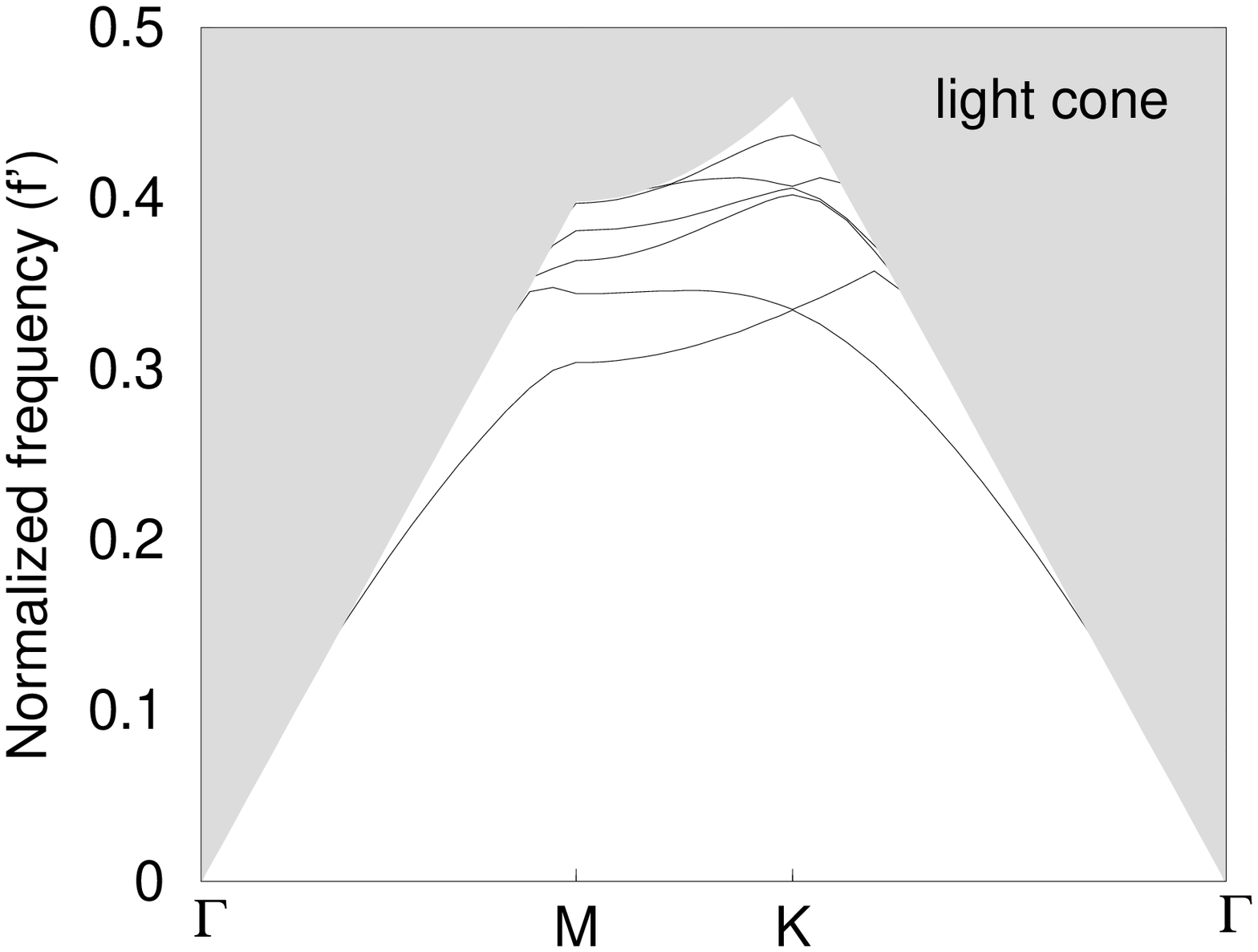,width=8cm}
\end{center}
\caption{ }
\label{fig:airsio2band}
\end{figure}


\newpage
\begin{references}
                                                                                
\bibitem{photonic_crystals} J. D. Joannopoulos, R. D. Meade and J. N. Winn,
 {\it Photonic Crystals} (Princeton University Press, Princeton, 1995)

\bibitem{photonic_band_gaps_and_localization} {\it Photonic Band Gaps and Localization}, 
 edited by C. M. Soukoulis (Plenum Press, 1993)

\bibitem{pre68_026704} R. L. Chern, C. C. Chang, C. C. Chang and R. R. Hwang,
 Phys. Rev. E {\bf 68}, 026704 (2003)

\bibitem{fan97} S. Fan, P. R. Villeneuve, J. D. Joannopoulos, 
 and E. F. Schubert, Phys. Rev. Lett. {\bf 78}, 3294 (2003).

\bibitem{yfchen} H. K. Fu, Y. F. Chen, R. L. Chern and C. C. Chang,
   Optics Express {\bf 13}, 7854 (2005).

\bibitem{op8_173} S. G. Johnson and J. D. Joannopoulos,
 Opt. Exps. {\bf 8}, 173 (2000); http://ab-initio.mit.edu/mpb/

\bibitem{roadmap} {\it Roadmap on Photonic Crystals}, edited by
 S. Noda and T. Baba (Kluwer Academic Publishers, 2003)

\bibitem{prb60_5751} S. G. Johnson, S. Fan, P. R. Villeneuve and J. D. Joannopoulos,
 Phys. Rev. B {\bf 60}, 5751 (1999)

\bibitem{pn1_1} C. Jamois, R. B. Wehrspohn, L. C. Andreani, C. Hermann, Ol Hess and U. Gosele,
 Photonics and Nanostructures -- Fundamentals and Applications {\bf 1}, 1 (2003)

\bibitem{prb66_033103} M. Qiu, Phys. Rev. B {\bf 66}, 33103 (2002)
                                                                                                       
\bibitem{ijqe38_891} L. C. Andreani and M. Agio, IEEE J. Quantum Electron. {\bf 38}, 891 (2002)

\bibitem{ijqe38_743} T. Baba, A. Motegi, T. Iwai, N. Fukaya, Y. Watanabe and A. Sakai,
 IEEE J. Quantum Electron. {\bf 38}, 743 (2002)

\end{references}
\end{document}